\documentclass[12pt,preprint]{aastex}

\shorttitle{Slow Diffusion of Cosmic-Rays}
\shortauthors{Fujita et al.}

\begin{document}

\title{Slow Diffusion of Cosmic-Rays around a Supernova Remnant}

\author{Yutaka Fujita, Yutaka Ohira, and Fumio Takahara}
\affil{Department of Earth and Space Science, Graduate School of
Science, Osaka University, 1-1 Machikaneyama-cho, Toyonaka, Osaka
560-0043, Japan}

\begin{abstract}
 We study the escape of cosmic-ray protons accelerated at a supernova
 remnant (SNR). We are interested in their propagation in interstellar
 medium (ISM) after they leave the shock neighborhood where they are
 accelerated, but when they are still near the SNR with their energy
 density higher than that in the average ISM. Using Monte-Carlo
 simulations, we found that the cosmic-rays with energies of
 $\lesssim$~TeV excite Alfv\'en waves around the SNR on a scale of the
 SNR itself if the ISM is highly ionized. Thus, even if the cosmic-rays
 can leave the shock, scattering by the waves prevents them from moving
 further away from the SNR. The cosmic-rays form a slowly expanding
 cosmic-ray bubble, and they spend a long time around the SNR.  This
 means that the cosmic-rays cannot actually escape from the SNR until a
 fairly late stage of the SNR evolution. This is consistent with some
 results of Fermi and H.E.S.S. observations.
\end{abstract}


\section{Introduction}
\label{sec:intro}

Most of the cosmic-rays in the Galaxy are believed to be accelerated at
the shock of supernova remnants (SNRs). Compared to cosmic-ray electrons
that can be directly observed \citep[e.g.][]{koy95}, observational
confirmation of accelerated protons is not easy. However, $\gamma$-ray
observations have suggested that the protons illuminate molecular clouds
around an SNR and produce $\gamma$-ray emission through $pp$-interaction
\citep[e.g.][]{aha04,aha08b,abd09a}. Hereafter, we treat cosmic-ray
protons.

Recent $\gamma$-ray observations of molecular clouds around an SNR have
given us information not only on the particle acceleration in the SNR
but also on the escape of the particles from the SNR. By comparing a
simple model with latest Fermi and H.E.S.S. $\gamma$-ray observations,
\citet{fuj09a} indicated that there are TeV cosmic-rays around the
hidden SNR in the open cluster Westerlund~2, and the old SNR W~28. They
showed that the diffusion time-scale of cosmic-rays around the SNRs is
much longer than that in the general region in the Galaxy. The diffusion
coefficient is less than $\sim 1$~\% of that in the general region
\citep[see also][]{tor08}. This suggests an important fact. The typical
diffusion coefficient in the general region is $D\sim 10^{29}\rm\:
cm^2\: s^{-1}$ for particles with an energy of $\sim 1$~TeV
\citep{ber90}. If the coefficient is 1\% of that typical value or $D\sim
10^{27}\rm\: cm^2\: s^{-1}$, the time-scale on which the particles cross
the scale of an SNR ($R_{\rm sh}\sim 15$~pc) is $\sim R_{\rm sh}^2/(6
D)\sim 1\times 10^4$~yr. This is comparable to the time in which a shock
with a velocity of $\sim 10^3\rm\: km\: s^{-1}$ crosses that scale. This
means that particles that can leave the shock neighborhood would not
easily escape from the periphery of the SNR. Here, the shock
neighborhood means the region where some non-linear effects generate
strong magnetic waves or cause strong amplification of magnetic fields
\citep{luc00,bel04}. In that region, the particle diffusion would follow
the so-called Bohm diffusion and the particles are efficiently
accelerated.

In this letter, we theoretically investigate particle diffusion in the
ISM outside this shock neighborhood. We show that cosmic-ray streaming
generates Alfv\'en waves around the SNR, which scatter the particles and
make the particle diffusion in the ISM around the SNR significantly
slow. We also show that the cosmic-rays actually cannot easily escape
from the periphery of the SNR. In contrast with previous studies
(e.g. \citealt*{ptu03}; see also \citealt*{lee08}), we consider the
time-evolution of the diffusion coefficient and the effect of cosmic-ray
diffusion in the ISM before being affected by the streaming. We use
Monte-Carlo simulations to study the diffusion, while at the same time
we calculate the evolution of the SNR and the excitation of Alfv\'en
waves based on a simple model. We consider the diffusion in the ISM
around an SNR that are highly ionized, because ion neutral friction does
not significantly affect the growth of the Alfv\'en waves. Observed
molecular clouds that are bright in the $\gamma$-ray band may be
immersed in the ionized ISM. In \citet{fuj09a}, it was not specified
whether the observed slow diffusion of cosmic-rays is happening inside
or outside the molecular clouds. In this study, we take the latter
stance.

\section{Models}

Cosmic-rays are scattered by Alfv\'en waves generated by the streaming
of the cosmic-rays themselves. We treat the propagation of a cosmic-ray
particle as a three-dimensional random walk. We assume that the
diffusion coefficient depends on the position $\mbox{\boldmath{$r$}}$,
although the scattering is isotropic at each point. For a given
diffusion coefficient $D(t,\mbox{\boldmath{$r$}}, E)$, where $E$ is the
energy of the particle, the mean free path of a particle is written as
$l_m=6 D/v$, where $v$ is the velocity of the particle. The diffusion
coefficient is given by
\begin{equation}
\label{eq:Da}
 D_a=\frac{4}{3\pi}\frac{p v c}{e B \psi}\:,
\end{equation}
where $p$ is the momentum of the particle, $c$ is the speed of light,
$e$ is the positron charge, $B$ is the magnetic field, and
$\psi(t,\mbox{\boldmath{$r$}}, E)$ is the energy density of Alfv\'en
waves per unit logarithmic bandwidth (which are resonant with particles
of energy $E$) relative to the ambient magnetic energy density $U_{\rm
M}$ \citep{bel78}. At the rest frame and the outside of a shock,
\begin{equation}
\label{eq:psi}
 \frac{\partial \psi}{\partial t} \approx \sigma - \Gamma\psi\:,
\end{equation}
where $\Gamma$ is the damping rate \citep{bel78}. The growing term is
given by
\begin{equation}
\label{eq:sigma}
 \sigma(t,\mbox{\boldmath{$r$}}, E) 
= \frac{4\pi}{3}\frac{v_{\rm A} p^4 v}{U_{\rm M}}
|\nabla f|\:,
\end{equation}
where $v_{\rm A}$ is the Alfv\'en velocity, and
$f(t,\mbox{\boldmath{$r$}}, E)$ is the distribution function of
cosmic-rays \citep{ski75,bel78}. For the damping, we consider the one
owing to collisions between charged and neutral particles;
\begin{equation}
\label{eq:gamma}
 \Gamma = 8.4\times 10^{-9} 
\left(\frac{n_{\rm H}}{\rm cm^{-3}}\right)
\left(\frac{T}{10^4\rm\: K}\right)^{0.4}
\min\left(1,\frac{p_{\rm coh}^2}{p^2}\right)
\: s^{-1}\:,
\end{equation}
where $n_{\rm H}$ is the neutral hydrogen density, $T$ is the
temperature of the ISM, and 
\begin{equation}
\label{eq:pcoh}
 p_{\rm coh}c = 8 
\left(\frac{n_{\rm H}}{\rm cm^{-3}}\right)^{-3/2}
\left(\frac{T}{10^4\rm\: K}\right)^{-0.4}
\left(\frac{B}{1\rm\: \mu\: G}\right)^2 \rm\: GeV
\end{equation}
\citep{kul71,dru96}. For $p>p_{\rm coh}$, neutrals participate in the
coherent oscillations of the ions in the Alfv\'en wave, which decreases
the damping rate.

The evolution of an SNR follows the Sedov solution for $t_{\rm
sed}<t<t_{\rm rad}$;
\begin{equation}
 R_{\rm sh}(t)=R_{\rm rad}(t/t_{\rm rad})^{2/5}\:,
\end{equation}
where $t_{\rm sed}$ is the starting time of the Sedov phase, and $R_{\rm
rad}$ is the radius of the SNR when its radiative cooling becomes
effective at $t=t_{\rm rad}$. For particle acceleration in the shock
neighborhood, we adopt a simple model. We assume that only particles
with a maximum energy $E_{\rm max}(t)$ can escape from the shock at a
given time \citep*[see][]{ptu05,ohi09}. We simply assume that $E_{\rm
max}(t)\propto t^{-\delta}$, where $\delta$ ($>0$) is the parameter
\citep*{gab09}. Since we are mainly interested in the particle diffusion
outside the shock, we assume that the growth of Alfv\'en waves stops
inside the shock for simplicity ($\dot{\psi}=0$).

We assume that $\psi=0$ at $t=0$. When $\psi$ is close to zero, the
diffusion coefficient in equation~(\ref{eq:Da}) diverges. Thus, we use a
diffusion coefficient of $D=\min(D_a, D_s)$ in actual calculations,
where $D_s$ is the standard value in the Galaxy at positions far away
from SNRs:
\begin{equation}
\label{eq:Ds}
 D_s=10^{28}\left(\frac{E}{10\rm\: GeV}\right)^{0.5}
\left(\frac{B}{3\:\mu\rm G}\right)^{-0.5}\rm\: cm^2\: s^{-1}
\end{equation}
\citep{gab09}. We do not consider the cooling of particles and the
reacceleration of particles that are caught up by the SNR.

\section{Results}
\label{sec:res}

First, we consider cosmic-ray diffusion in the ISM that is completely
ionized ($\Gamma=0$).  We assume that $R_{\rm rad}=15$~pc and $t_{\rm
rad}=1\times 10^4$~yr, which are close to those for Westerlund~2 and
W~28 \citep{fuj09a,fuj09b}. The strength of undisturbed magnetic field
is $B=10\:\mu$~G. The mass density of interstellar medium (ISM) is
$\rho=8.35\times 10^{-24}\rm\: g\: cm^{-3}$. The maximum energy $E_{\rm
max}$ decreases from 100~TeV ($t=t_{\rm sed}=215$~yr) to 1~GeV
($t=t_{\rm rad}$), which leads to $\delta=3$. For the calculations of
$\psi$ (equation~[\ref{eq:psi}]), we set the inner boundary at
$r=3.2$~pc, where the highest energy cosmic-rays are injected
($E=100$~TeV). We consider the gradient of $f$ only in the radial
direction. The outer boundary is set at $r=50$~pc. The innermost mesh
has a width of $\sim 0.2$~pc, while the width of the outermost mesh is
$\sim 6$~pc. The time-step for the calculation of $\psi$ is taken as the
time in which $E_{\max}(t)$ decreases by a factor of $10^{0.2}$, which
is a factor of a few smaller than the duration in which particles with
$E_{\max}(t)$ are actually injected \citep*{cap09}.

In Fig.~\ref{fig:dist}, we present the distribution of particles with an
energy of $E=1$~TeV at $t=2.2\times 10^3$ and $1\times 10^4$~yr (or
$t_{\rm rad}$). The shock radius at those times is $R_{\rm sh}=8.1$, and
15~pc, respectively. The particles with that energy are injected at
$t=1\times 10^3$~yr at $R_{\rm sh}=6.0$~pc. The number of the injected
particles with that energy is 100,000. Immediately after the injection
at $t=1\times 10^3$~yr, the particles rapidly leave the shock because of
the large diffusion coefficient ($D=D_s$), until Alfv\'en waves grow
enough to scatter the particles.

Figure~\ref{fig:dist} indicates that many of the particles are still
wandering around the SNR at $t=2.2\times 10^3$~yr. Until this time, the
streaming of the cosmic-rays has generated Alfv\'en waves outside the
SNR, and they are strong enough to scatter particles (but still $\psi\ll
1$). As a result, the diffusion coefficient has become much smaller than
$D_s$ in equation~(\ref{eq:Ds}). In fact, at $r\sim 10$~pc, $D_a/D_s\sim
0.01$ for $E=1$~TeV, which is consistent with the Fermi and
H.E.S.S. results \citep{aha07,aha08b,abd09b} as is discussed in
\citet{fuj09a}. Even at $t=t_{\rm rad}$, a significant fraction of the
particles remain around the expanding SNR. Because of the relatively
high value of $\psi$ and the low value of the diffusion coefficient,
cosmic-rays cannot easily escape from the region peripheral to the SNR.
At $t=t_{\rm rad}$, 71\% of the injected particles with $E=1$~TeV are
around the SNR ($r<20$~pc). The number density does not decrease
smoothly toward larger $r$ (Fig.~\ref{fig:dist}), because the diffusion
becomes slow where the gradient of $f$ is large
(equation~[\ref{eq:sigma}]), which tends to steepen the gradient
further.

Figure~\ref{fig:spec} shows the spectrum of cosmic-rays at $t=t_{\rm
rad}$. We assume that the total energy of cosmic-rays accelerated for
$t_{\rm sed}<t<t_{\rm rad}$ is $10^{50}$~erg. Although we give the same
number of particles for a given energy in the calculations, we give them
weights so that the distribution function of all the particles that are
accelerated at the shock is $f_0\propto E^{-2}$. Figure~\ref{fig:spec}
indicates that $f\propto E^{-2.75}$ for $E\gtrsim 10$~TeV regardless of
positions. The index ($-2.75$) is the one when particles have already
prevailed beyond the region considered \citep{aha96,fuj09a}. For this
energy range, Alfv\'en waves are not much excited, because equation
(\ref{eq:Ds}) indicates that the diffusion coefficient when $\psi\approx
0$ ($D=D_s$) is large, and because the particles rapidly escape from the
SNR without making a gradient of $f$ large enough to excite Alfv\'en
waves (equation~[\ref{eq:sigma}]). On the other hand, the diffusion
coefficient when $\psi\approx 0$ is smaller for $E\lesssim 3$~TeV
(equation~[\ref{eq:Ds}]), Alfv\'en waves are generated through the
gradient of $f$ in this energy range. As a result, cosmic-rays in this
energy range are well scattered by the Alfv\'en waves, and spread only
slowly as a cosmic ray bubble. Thus, the spectrum follows the original
form of $f\propto E^{-2}$, and the value of $E^2 f$ is distinctively
large at $r\lesssim R_{\rm sh}$ ($=15$~pc) compared to the one at
$r>R_{\rm sh}$. The position of the cutoff in the spectra ($\sim 3$~TeV)
depends on the diffusion coefficient of the unperturbed ISM
(equation~[\ref{eq:Ds}]). This indicates the need to study the escape of
cosmic-rays globally, not locally around the shock.  At $r=32$--36~pc,
the spectrum is harder than $E^{-2}$ at $E\lesssim 3$~TeV, because most
of the low-energy particles have not reached that region. It is to be
noted that the wave energy does not exceed the background magnetic
energy around the SNR in this energy range ($\psi < 1$). This is
different from the shock neighborhood mentioned in \S~\ref{sec:intro},
where $\psi\gtrsim 1$ is expected. The retained cosmic-rays may be
released at $t\gtrsim t_{\rm rad}$ when the SNR can no longer hold them.

The spectra are different when the ISM is not completely ionized. In
Fig.~\ref{fig:spec2}, we assume that the ISM temperature is $T=10^4$~K,
and the neutral hydrogen density is $n_{\rm H}=0.1\rm\: cm^{-3}$
(equations~[\ref{eq:gamma}] and [\ref{eq:pcoh}]). We chose these
parameters because the neutral damping is marginally effective, and we
can see the effect clearly. Other parameters are the same as those
studied above. In Fig.~\ref{fig:spec2}, the neutral damping is effective
for $E<1$~TeV at $r\gtrsim 16$~pc. Around the SNR or $r\sim 12$--16~pc,
the generation of Alfv\'en waves overwhelms the damping at $E\sim 1$~TeV
because of the large gradient of $f$. At $r\sim 12$--16~pc, particles
that are recently injected are seen at $E\lesssim 10$~GeV.

\section{Summary and Discussion}

We have shown that cosmic-rays with energies of $\lesssim$~TeV excite
Alfv\'en waves in the ISM around an SNR even after they leave the shock
neighborhood where they are accelerated. The particles are scattered by
the waves, which makes the diffusion of the particles slow. As a result,
the particles remain around the expanding SNR for a long time.  This
means that even if the particles are accelerated in an early stage of
the SNR evolution, they cannot actually escape from the SNR until a
fairly late stage.

\begin{figure}
\epsscale{1.} \plotone{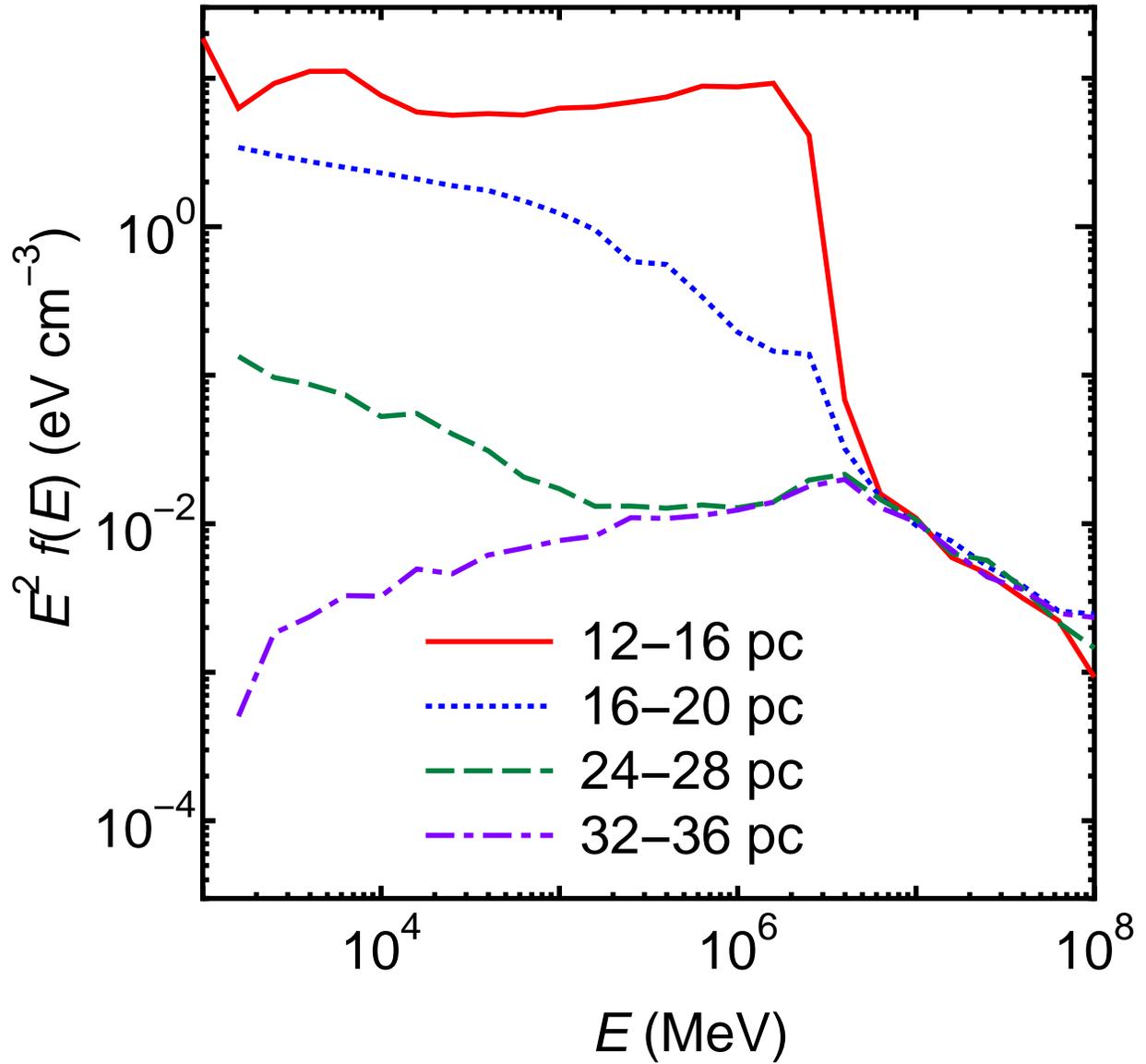} \caption{Energy spectra of cosmic-rays at
 $t=t_{\rm rad}=1\times 10^4$~yr. The ISM is fully ionized. The solid
 line is for $r=$12--16~pc, the dotted line is for 16--20~pc, the dashed
 line is for 24--28~pc, and the dot-dashed line is for
 32--36~pc. \label{fig:spec}}
\end{figure}

\begin{figure}
\epsscale{1.} \plotone{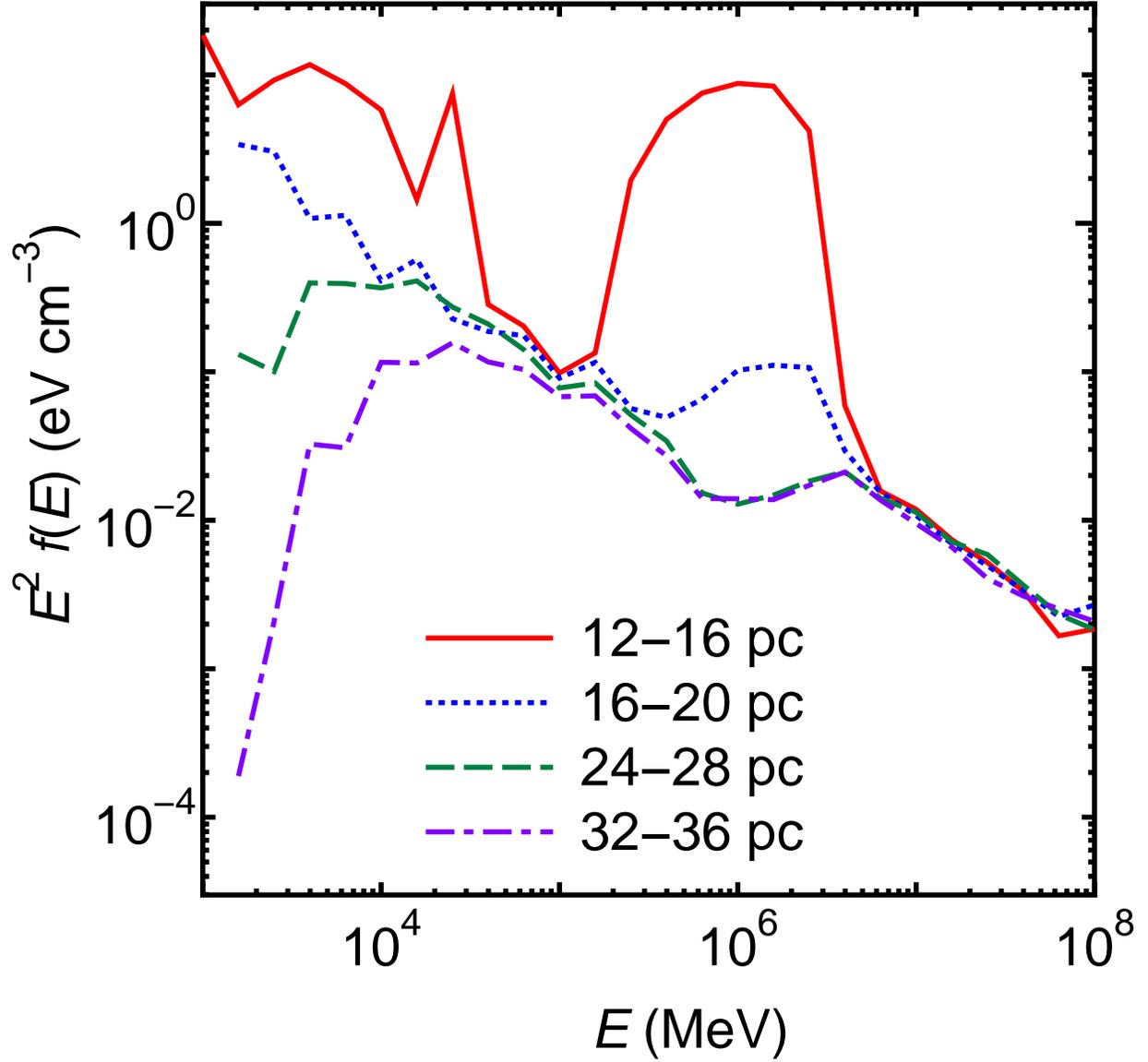} \caption{Same as Fig.~\ref{fig:spec} but
 the ISM is not completely ionized. \label{fig:spec2}}
\end{figure}

The ISM must have been well ionized for the cosmic-rays to be held
around the SNR, because the neutral damping of Alfv\'en waves is not
effective. For Westerlund~2 and W~28, \citet{fuj09a} indicated that the
supernova exploded in a hot cavity, which had been created through
activities of the progenitor star and/or explosions of other
supernovae. Therefore, the ISM might have been ionized. While the
maximum energy of protons around the SNR is $\sim 2.7$~TeV for W~28, it
is $\sim 47$~TeV for Westerlund~2 \citep{fuj09a}, which is larger than
that predicted in Fig.~\ref{fig:spec} ($\sim 3$~TeV). However, the
maximum energy increases to $\gtrsim 10$~TeV if the magnetic fields in
the ISM is $B\sim 20\:\mu$G and the ISM is fully ionized.

\begin{figure}
\epsscale{1.} \plotone{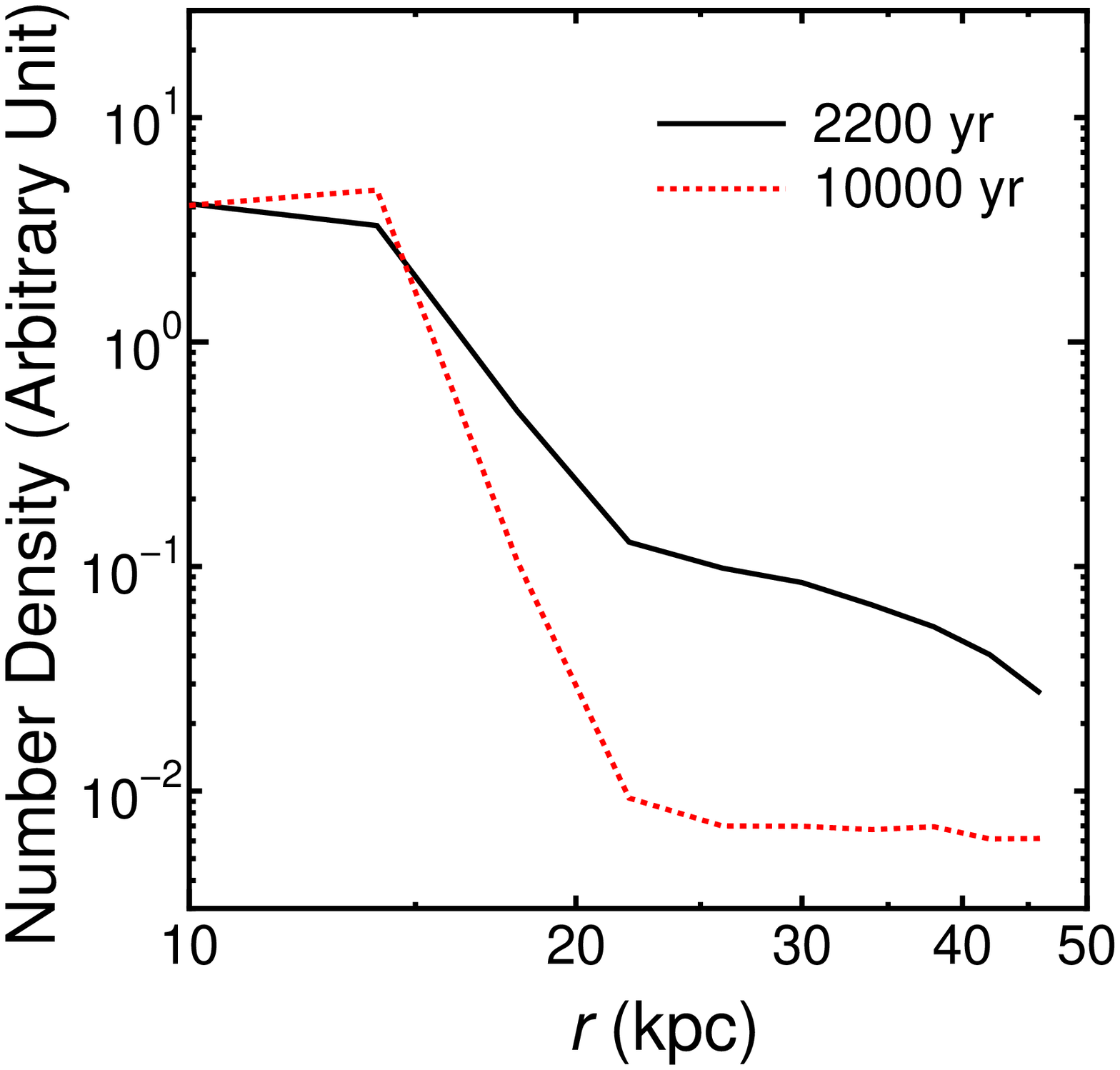} \caption{Number density of particles with
 an energy of 1~TeV at $t=2.2\times 10^3$ (solid) and $1\times 10^4$ yr
 (dotted). The radius of the SNR at those times is $R_{\rm sh}=8.1$ and
 15~pc, respectively. The supernova exploded at the
 origin. \label{fig:dist}}
\end{figure}

Recent H.E.S.S and Fermi observations showed that $\gamma$-rays are
produced in molecular clouds adjacent to SNRs
\citep[e.g.][]{aha04,aha08b,abd09a}. In molecular clouds, Alfv\'en waves
may totally dissappear through neutral damping. If this is the case,
they may be illuminated by cosmic-rays in the ionized ISM surrounding
the clouds. The $\gamma$-rays may be generated through $pp$-interaction
in the clouds. The cosmic-rays that enter the clouds may quickly pass
the clouds unless some other disturbers such as turbulence scatter the
cosmic-rays. In our study, the cosmic-rays are confined around an SNR on
a scale of the SNR itself (Fig.~\ref{fig:dist}). Thus, molecular clouds
that are bright in the $\gamma$-ray band may not be found in the region
beyond this scale.

If cosmic-rays continue to be held by an SNR, they may be affected by
adiabatic cooling as the SNR expands. Moreover, some of the cosmic-rays
engulfed by the SNR would be reaccelerated. These may affect the
cosmic-ray spectrum in the Galaxy.

\acknowledgments

This work was supported by KAKENHI (Y.~F.: 20540269; Y.~O.: 20.1697;
F.~T.; 20540231).

\clearpage

\clearpage

\end{document}